\documentstyle[manuscript,aps]{revtex}
\begin{document} 
\draft
\title{Possible solution to basic problems 
regarding the coupling constants $G_V$ and $G_A$}
\author{V. P. Gudkov} 
 \address{
Department of Physics and Astronomy,  
University of South Carolina,   
Columbia, SC 29208}
\author{ K. Kubodera}
 \address{
Department of Physics and Astronomy,  
University of South Carolina, 
Columbia, SC 29208\footnote{permanent address}\\
and CEA/Saclay, 
Service de Physique Th{\' e}orique\\
F-91191, Gif-sur-Yvette Cedex, France}
\maketitle 
\begin{abstract}
The existing experimental evidence
on the vector and axial-vector coupling constants,
$G_V$ and $G_A$, for neutron $\beta$-decay exhibits 
two prominent problems:
(i) the unitarity for the first row of the CKM matrix 
seems violated;
(ii) one obtains different values of $G_A/G_V$ 
according to whether one uses as input the neutron lifetime 
or neutron-decay correlation observables.
We show that the in-medium modification of $G_A$
can influence both the inner and outer radiative corrections
used in deducing $G_V$ 
from the super-allowed Fermi transitions
and that this effect may resolve the long-standing problems 
(i) and (ii) simultaneously.
\end{abstract}


\newpage

The vector coupling constant $G_V$ 
and the axial-vector coupling constant $G_A$
for neutron $\beta$-decay
are quantities of primary importance
that have been studied extensively.
The recent remarkable progress 
in the determination of $G_V$ and $G_A$
highlights two very notable problems.
One problem is concerned with $G_V$ and 
the Cabibbo-Kobayashi-Maskawa (CKM) matrix element 
$V_{ud}$.
Comparison of $G_V$ with 
$G_\mu$ obtained from the $\mu$-decay rate 
gives the empirical value of $V_{ud}$,
and the existing data give \cite{towst}
\begin{equation}
V_{ud}= 0.9740\pm 0.0005.
\label{Vudnow}
\end{equation}
Since in the standard model the CKM matrix is unitary,
the first row of the CKM matrix must satisfy
$ |V_{ud}|^2 +|V_{us}|^2+|V_{ub}|^2= 1$.
However, $V_{ud}$ in Eq.(\ref{Vudnow})
combined with the available information on 
$V_{us}$ and $V_{ub}$ leads to
\begin{equation}
|V_{ud}|^2 +|V_{us}|^2+|V_{ub}|^2= 
0.9972 \pm 0.0013, \label{unito}
\end{equation}
indicating a  violation of unitarity at the $98\%$ 
 confidence level\cite{towst}.
Since this violation, if confirmed,
constitutes a clear signal for physics beyond the standard model,
and since the contribution of $V_{ud}$ dominates
the sum in Eq.(\ref{unito}),
it is vitally important to assess the reliability 
of $V_{ud}$ in Eq.(\ref{Vudnow}). \par
The second problem concerns $\lambda \equiv G_A/G_V$.
The neutron-decay angular distribution experiments give
\cite{pdg}
\begin{equation}
\lambda _A\,=\,1.2601 \pm 0.0025,
\label{GAGV1}
\end{equation} 
whereas the neutron lifetime measurements,
which have recently greatly improved, 
give \cite{pdg} 
\begin{equation}
\lambda _{\tau }\,=\,1.2681\pm0.0033.
\label{GAGV2}
\end{equation}
(For the neutron lifetime we  use the value $888\pm 3$ sec 
which is between  the two recommended values 
$887\pm 2.0$ sec and $889.2\pm 2.2$ sec.) 
The origin of this discrepancy in $\lambda$ 
is at present unknown,
a situation that is rather disturbing. \par
In this Letter we point out the possibility
that the above two problems are related to each other
and that they may be solved simultaneously
by taking into account  certain types of hadronic effects
which have not been considered so far.
As is well known,
$G_V$ is obtained from the observed $ft$ values
for $0^+\rightarrow 0^+$ nuclear $\beta$-decays
after applying radiative corrections 
and isospin-mixing corrections.
Although these corrections have been studied
in great detail by many authors,
we present here a new type of nuclear correction
 which increases $G_V$, 
and hence $V_{ud}$ as well, in such a way 
that the unitarity condition is satisfied.
This increase in $G_V$ decreases the value of 
$\lambda_{\tau}$ due to the fact 
that the neutron decay rate is proportional to 
$ G_V^2+3G_A^2\,=\, G_V^2(1+3\lambda_{\tau}^2)$.
By contrast, $\lambda_A$ remains unaffected,
because it is directly related to the correlation observables
in neutron $\beta$-decay and hence has no connection
with the $ft$ values of the $0^+\rightarrow 0^+$ transitions.
We will argue that our new value of $\lambda_{\tau}$
is compatible with $\lambda_A$. 

In deducing the empirical value of $V_{ud}$
from comparison of the rates of 
$0^+\rightarrow 0^+$ nuclear $\beta$-decays 
and the strength of the pure leptonic $\mu $-decay,  
an accurate estimate of 
radiative corrections is the most important step
(see for example,\cite{towst,sirl,marc,hardb} 
and references therein). 
The bench mark work which constitutes 
the basis of all the recent analyses
is due to Marciano and Sirlin \cite{marc}, 
who find the following corrections
\begin{equation}
\Delta  _{O(\alpha )} = 
{\alpha \over{2 \pi }} \left[\, \overline{g}(E_m)
+ 4 \ln{(m_Z / M)} +  
\ln{(M / M_A)} + 2 C +{\it A_g}
\right] . \label{corr}
\end{equation}
The first term, often called the outer correction,
represents a spectrum-average effect which depends on 
the $\beta$-ray end-point energy $E_m$,
and its contribution amounts to $\sim0.01$\cite{jaus}. 
Of the remaining terms (inner corrections),
the $4\ln{(m_Z / M)}$ term represents 
the dominant model-independent short-distance contribution 
($M$ = proton mass, $m_Z$ = Z-boson mass),
whereas $\ln{(M / M_A)} + 2 C$
are axial-current induced contributions.
$M_A$($\sim$1 GeV) is a low-energy cutoff characterizing 
the short-distance part of the $\gamma W$ box diagram,
while $C$ represents the long-distance correction.
Numerically, the contributions of 
the short-distance and long-distance parts are comparable,
each contributing about $ 0.001$ (see, e.g. ref.\cite{tow94}).
Finally, ${\it A_g}$ is a perturbative QCD correction,
which turns out to be very small.
The standard radiative corrections above 
have been obtained for the weak quark current
$\overline{q}\gamma ^{\mu }(1 + \gamma _5) q$.
In applying theses results to nucleon 
and nuclear $\beta$-decays,
one needs to translate the quark-based description 
to hadronic descriptions,
and this step is considered to introduce
a certain amount of model dependence.

The fact that the axial current can contribute to
the radiative correction of the Fermi transition means
that, if one uses the nucleon weak current 
instead of the quark current,
the results will in general depend on $\lambda$,
the strength of the axial-current coupling,
and that the well-established 
nuclear-medium dependence of $\lambda$
is capable of affecting the radiative correction.
In fact, parts of such effects
have been studied in the literature.
Several authors \cite{jaus,tow92,bark} investigated
the two-nucleon contribution to the $C$ term
in Eq. (\ref{corr}),
while Towner \cite{tow94} estimated the influence 
of the in-medium modification of $\lambda$ 
on the $C$ term.
Changes in $C$ due to these effects 
were reported to be small.
The existing work is limited to the $C$ term because
in the treatment of Ref.\cite{marc},
the $\ln{(M / M_A)}$ term,
representing the short-distance effects
was considered to be independent of 
hadron/nuclear models,
leaving only the $C$ term 
as a model-dependent contribution.
We note, however, that, although the separation
of the axial-vector induced corrections
into short- and long-distance parts
has a clear-cut meaning in the quark picture,
there is a possibility that this separation 
becomes less well defined 
when one switches (as one must do at some point) 
to hadronic descriptions. 

In this note we estimate 
the effects of the in-medium modification of $\lambda$, 
using the nucleonic weak current from the outset,
and without invoking separation 
into short- and long-distance contributions.
This means that our calculation in principle 
incorporates the combined effects 
of the $\ln(M/M_A)$ and $C$ terms, 
even though we do not try to make direct contact 
with the quark-model-based classification 
used in Eq. (\ref{corr}).
As will be demonstrated below,
our approach indeed gives rise 
to a cut-off dependent term which is 
a function of $\lambda$ and hence can be influenced 
by nuclear-medium effects.
It is important to emphasize, however,
that we work here with 
the nucleonic weak current instead of 
the quark weak current {\it not} because
we think the former has more fundamental meaning.
Rather, the reason is that 
the nucleonic current description provides
an interesting alternative framework
for incorporating the nuclear medium effects.

Our work has some overlap 
with that of Towner \cite{tow94}
in that the effects of in-medium modification
of $\lambda$ are considered
but, as mentioned above, the calculation
in \cite{tow94} was limited to the $C$ term.
Obviously, a calculation of the 
entire $\lambda$-dependent effect 
(rather than just the $C$ term)
is an extremely difficult task. 
In the present exploratory work, therefore,
we present an estimate based on schematic models.
Thus, we use the nucleon weak current
\begin{equation}
\overline{u}_p\gamma ^{\mu }
(1 + \lambda \gamma _5) u_n ,
\end{equation}
and assume that the nuclear-medium effect 
can be represented by changing the free space value 
$\lambda$ into an in-medium value $\lambda^*$. 
As for electromagnetic interactions 
we take into account 
the nucleon charge and magnetic moments
and  the electron charge. 
To calculate the loop diagrams, 
we employ two simplified approaches.  
In the first approach 
we use a single common cut-off parameter 
$\Lambda _0$ instead of form factors at the vertices. 
In the second approach we use a common form factor
$\Lambda ^2/(q^2 +\Lambda ^2)$,
both at the weak and at the electromagnetic vertices, 
where  $q$ is transferred momentum, 
and $\Lambda$ is the parameter 
of the same order of magnitude
as the nucleon mass $M$. 

For each of the two cases, 
a simple but quite lengthy calculation 
leads to the following expression 
for a $\lambda$-dependent contribution 
to the inner radiative correction: \\
Case I:
\begin{equation}
\Delta_{inner} =  \lambda {\alpha \over{2 \pi }}  
{1\over{{\xi}_0^2}} \left(
(\overline{\mu}_p - \overline{\mu}_n ) 
( {\xi}_0 -1) \ln{
(1-{\xi}_0 ) }+ 2 [{\xi}_0 -2 - \ln{(1 - {\xi}_0 )} ] \right) ,  
\label{delone}
\end{equation}
Case II:
\begin{equation}
\Delta _{inner} =  \lambda {\alpha \over{2 \pi }}  \left(
(\overline{\mu}_p - \overline{\mu}_n ) 
({{13}\over{12}} - \xi \ln{\xi}) + 
{7\over{6}} + 2 \xi \ln{\xi} 
+ {1\over{\xi ^2}} + {3\over{4}}\ln{\xi } - 
{1\over{4 \xi }} \ln{\xi } \right) ,  \label{deltwo}
\end{equation}
where $\overline{\mu}_p$ and $\overline{\mu}_n$ 
are the proton and neutron anomalous magnetic moments, 
${\xi}_0 =  M^2 / {\Lambda }_0 ^2$ and  
$\xi = M^2 / \Lambda ^2$.  
We have checked 
that the two rather different 
phenomenological approaches used here 
give numerically similar results
within reasonable variations of the model parameters,
$\Lambda _0$ and $\Lambda$.  
 Even in the absence of these cut-off parameters 
(point-like nucleons with a cut-off at the scale of the vector 
 boson mass) the calculated inner corrections does not
 exceed the values in Eqs.(\ref{delone}) and (\ref{deltwo})
by more than an order of magnitude. 
This indicates a certain degree of stability of our results.  

In the above we concentrated on the inner corrections.
We now turn our attention to the outer correction,
which is defined as an $E_m$-dependent correction.
As is well known\cite{sirl},
if one only retains terms of $O(\alpha)$,
ignoring terms of $O(\alpha E/M)$ 
and $O(\alpha(E/M)\ln(M/E))$ ($E$= electron energy),
then the outer correction becomes
model-independent and hence also 
$\lambda$-independent.
The term $\overline{g}(E_m)$ in Eq. (\ref{corr})
represents this model-independent contribution.
However, at the level of precision 
we are interested in,
it becomes necessary to retain terms of 
$O(\alpha E/M)$ and $O(\alpha(E/M)\ln(M/E))$,
and these terms lead to 
a $\lambda$-dependent outer correction.
We write a generic form 
of the $\lambda$-dependent outer correction as
\begin{equation}
\Delta_ {outer} =  
\lambda\,{\alpha \over{2 \pi }}\,f(E_m ).
 \label{delout}
 \end{equation}
The explicit forms of $f(E_m )$, 
which depend on the models used,
are very lengthy and will be given elsewhere.
We remark that the separation 
of the radiative corrections 
into the terms in Eq.(\ref{corr}) is specific 
to the quark model,
and therefore it is not possible to establish 
termwise correspondence between 
the terms in Eq.(\ref{corr}) and those in 
Eqs.(\ref{delone}) - (\ref{delout}).

We denote by $\Delta_\lambda$ 
the leading-order $\lambda$-dependent radiative correction,
which is the sum of the inner correction in Eq. (\ref{delone}) 
or Eq. (\ref{deltwo}),
and the outer correction in Eq. (\ref{delout}):
\begin{equation}
\Delta_ {\lambda } \equiv  \Delta _{inner} 
+ \Delta_ {outer} 
\equiv\, \lambda \cdot {\alpha \over{2 \pi }} J.   
\label{della}
\end{equation}
Here $J $ contains information about 
the nucleon structure including
its weak and electromagnetic form factors. 
Assuming that the cut-off parameters  
in Eqs.(\ref{delone})-(\ref{delout}) 
are controlled by a strong interactions scale 
($\Lambda_0 \sim \Lambda \sim M$) and 
that a characteristic energy of nuclear $\beta$-decay 
is $E_m \sim 5$ MeV, 
the value of $J$ turns out to be typically $5 \sim 25$.  
In the subsequent numerics, we use $J=15\pm10$.

Since we have a general expression 
(within our schematic model) 
for the $\lambda$-dependent part 
of the $O(\alpha)$ radiative corrections,
we may use it to study the influences 
of the in-medium modification of $\lambda$.
We expect that, if a description based on
the nucleon weak current achieves a sufficient level
of precision, then the result corresponding to
the case with $\lambda$ {\it fixed at the free nucleon value}
should essentially reproduce the results of
the existing analyses of the relevant observables
based on Eq.(\ref{corr}).
Then we may assume that small perturbative changes 
in the $\lambda$-dependent radiative correction 
due to in-medium modification of $\lambda$
can be estimated with the present schematic model. 
We use Eq.(\ref{della}) in this limited context.

We introduce the in-medium renormalization parameter 
$\rho$ as $\rho \equiv \lambda^*/\lambda$.
For a wide range of the periodic table,
a typical value of $\rho$ is $\rho \sim 0.8$ \cite{alk,hard}.
For our present purpose, we fix $\rho$ at 0.8.
Then, by taking the difference of $\Delta_{\lambda}$, 
Eq. (\ref{della}), between the nuclear and free cases,
we obtain
\begin{equation}
\Delta \equiv  \Delta_ {\lambda} - \Delta_ {\lambda^*} =
\lambda\, (1-  \rho ) {\alpha \over{2 \pi }} J . \label{delt}
\end{equation}
This leads to a corrected value\cite{hardb} 
of the CKM matrix element $V_{ud}$ 
\begin{equation}
V_{ud} \rightarrow V_{ud} (1+ \Delta /2), \label{matr}
\end{equation}
and, since the probability of neutron $\beta$-decay 
is proportional 
to the $|V_{ud}|^2$, to a corrected expression 
for the neutron life time
\begin{equation}
\tau_n = {{2(Ft)}\over{f_R \ln 2}}
{{(1-\Delta )}\over{(1+3\lambda _{\tau}^2)}}. \label{neut}
\end{equation}
Here $F$ and $f_R$ are nuclear 
and neutron phase space factors\cite{hardb,wilk}. 
Eqs.(\ref{matr})  and  (\ref{neut}) summarize 
how the medium dependence of $\lambda$
encoded in $\Delta$ can change the deduction of $V_{ud}$ 
and the analysis of $\tau_n$.
For the choice of the values of parameters explained above,
we have $\Delta = 0.0044\pm 0.0029$. \par
The new formula Eq.(\ref{matr}) changes 
the value of $V_{ud}$ from
$V_{ud} = 0.9740 \pm 0.0005$ to
$V_{ud} = 0.9761 \pm 0.0015$.  
The error for the  CKM matrix element 
 has been estimated simply as the sum of the 
squared errors, one coming from Eq.(\ref{Vudnow}) and 
the other coming from the allowed range of $J$. 
The new value of the CKM matrix element in turn leads to 
\begin{equation}
|V_{ud}|^2 +|V_{us}|^2+|V_{ub}|^2= 1.0013 
\pm  0.0031\label{unitn}
\end{equation}
in good agreement with the unitarity condition.  
Moreover, the use of Eq. (\ref{neut}) 
with $\Delta=0.0044\pm 0.0029$
decreases $\lambda_{\tau}$  from the value in Eq.(\ref{GAGV2}) 
to $\lambda_{\tau}=1.2655\pm 0.0034$.
This new value agrees with $\lambda_A$, 
Eq.(\ref{GAGV1}) within the error bars. 

Thus, we have illustrated the possibility 
that the inclusion of the effect of the in-medium modification
of $G_A$ in the entire radiative correction 
(not only in the $C$ term)
is capable of removing the two outstanding problems
regarding the basic coupling constants $G_V$ and $G_A$.
Our present treatment is admittedly very schematic
and subject to various improvements,
and the particular numerical successes reported above
are simply indication of the possible effect.
However, we believe that
our calculation does indicates the importance 
of the new type of radiative correction considered here 
and that future analyses of the $0^+ \rightarrow 0^+$
$\beta$-transitions need to include the effects 
discussed in this article.

\acknowledgments
The authors are grateful to F. Myhrer and B.M. Preedom for useful discussions. 
KK wishes to thank J.N. Bahcall and W.C. Haxton,
who organized the INT Workshop on Solar Nuclear Fusion,
where he was able to learn the latest status of the problems discussed 
in this paper.  \par
This work is supported in part by the National Science Foundation, USA 
Grant No. PHYS- 9602000.

\end{document}